\documentclass[a4paper,11pt]{article}
\usepackage{pos}
\usepackage{graphicx}
\usepackage{subfig}

\title{VERITAS follow-up observation of the BL Lac blazar B2 1811+31 2020 Flare}
 \ShortTitle{VERITAS follow-up of BL Lac B2 1811+31 2020 Flare}

\author*[a]{Pablo Drake}
\author[a]{Colin Adams}
\onbehalf{for the VERITAS collaboration}

\affiliation[a]{Physics Department, Columbia University \\
  538 West 120th Street, 704 Pupin Hall, NY 10027, USA,}

\emailAdd{pd2629@columbia.edu}
\emailAdd{ca2762@columbia.edu}

\abstract{VERITAS is an imaging atmospheric Cherenkov telescope (IACT) array most sensitive to gamma rays in the very-high-energy (VHE) energy band (85 GeV - 30 TeV). As a part of its active galactic nuclei (AGN) program, VERITAS focuses on the identification and follow-up of AGN flares reported by other multiwavelength observatories. Between October 15th and October 19th, 2020, VERITAS followed up on the \textit{Fermi}-LAT and MAGIC detections of a flare of the intermediate-frequency-peaked BL Lacertae (IBL) object, B2 1811+31, located at a redshift of z=0.117. In this work, we present preliminary scientific results from the analysis of B2 1811+31’s 2020 flare, including the corresponding \textit{Fermi}-LAT light curve and VERITAS detection analysis.}

\ConferenceLogo{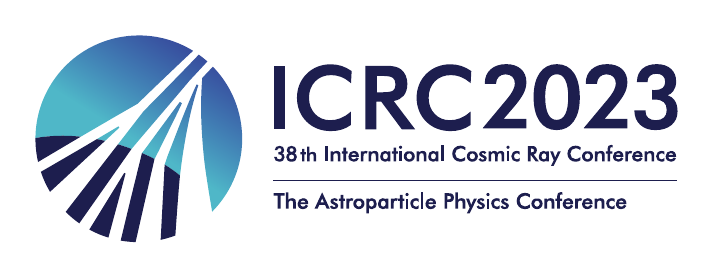}

\FullConference{%
38th International Cosmic Ray Conference (ICRC2023)\\
  26 July - 3 August, 2023\\
  Nagoya, Japan}

\begin{document}
\maketitle

\section{Introduction}
On 1 October 2020, the Large Area Telescope (LAT), one of the two instruments on the \textit{Fermi} Gamma-ray Space Telescope, measured an 11-factor flux increase in the daily averaged gamma-ray flux (E>100 MeV) of 4FGL J1813.5+3144 (referred to in this work as B2 1811+31), relative to the average flux reported in the fourth \textit{Fermi}-LAT catalog (4FGL). This event was sent out as an alert \cite{Angioni2020ATel} and prompted a multi-wavelength campaign from the optical band \cite{Bonnoli2020ATel} to very high energy (VHE, E > 100 GeV) gamma-rays. In fact, follow-up observations led to the first detection of this blazar in VHE by the MAGIC telescopes \cite{Blanch2020ATel}, reported on October 13th, 2020. Two days later, VERITAS, a ground-based gamma-ray detector sensitive to photons in the VHE, 85 GeV - 30 TeV range, started a 5-night campaign that observed the source from October 15th to October 19th \cite{Quinn2020ATel}. It resulted in a preliminary 7$\sigma$ detection, that, after our updated analysis, amounted to an  8.5\(\sigma\) detection of B2 1811+31 with 4.35 hours of observations. \\

\quad In this work, we characterize the evolution of the B2 1811+31 2020 flare with \textit{Fermi}-LAT to understand the parallel evolution of the source in HE and VHE wavelengths. We contextualize VERITAS's detection of the source within the longer evolution of the flare light curve.

\section{Observations}
\subsection{VERITAS}
VERITAS \cite{Holder2006AP}, the Very Energetic Radiation Imaging Telescope Array System, is a ground-based gamma-ray detector located at the Fred Lawrence Whipple Observatory (FLWO) in southern Arizona. The VERITAS array comprises four 12-meter imaging atmospheric Cherenkov telescopes.  Each telescope has a Davies-Cotton-design segmented mirror dish with 345 facets, and each dish is equipped with a 499 PMT camera, with a total field of view of 3.5$^\circ$. The 68\% containment radius for a 1 TeV photon is < 0.1$^\circ$, and the pointing accuracy is < 50". In its current configuration, VERITAS provides a 5$\sigma$ detection of a source with flux 1\% that of the Crab Nebula in about 25 hours of observations \cite{Park2015ICRC}. VERITAS data has been analyzed with the standard VERITAS software \texttt{VEGAS} \cite{Cogan2008ICRC}.

\begin{figure}[!tbp]
  \centering\includegraphics[width=0.65 \textwidth, trim={0 0 0 2.1cm}, clip]{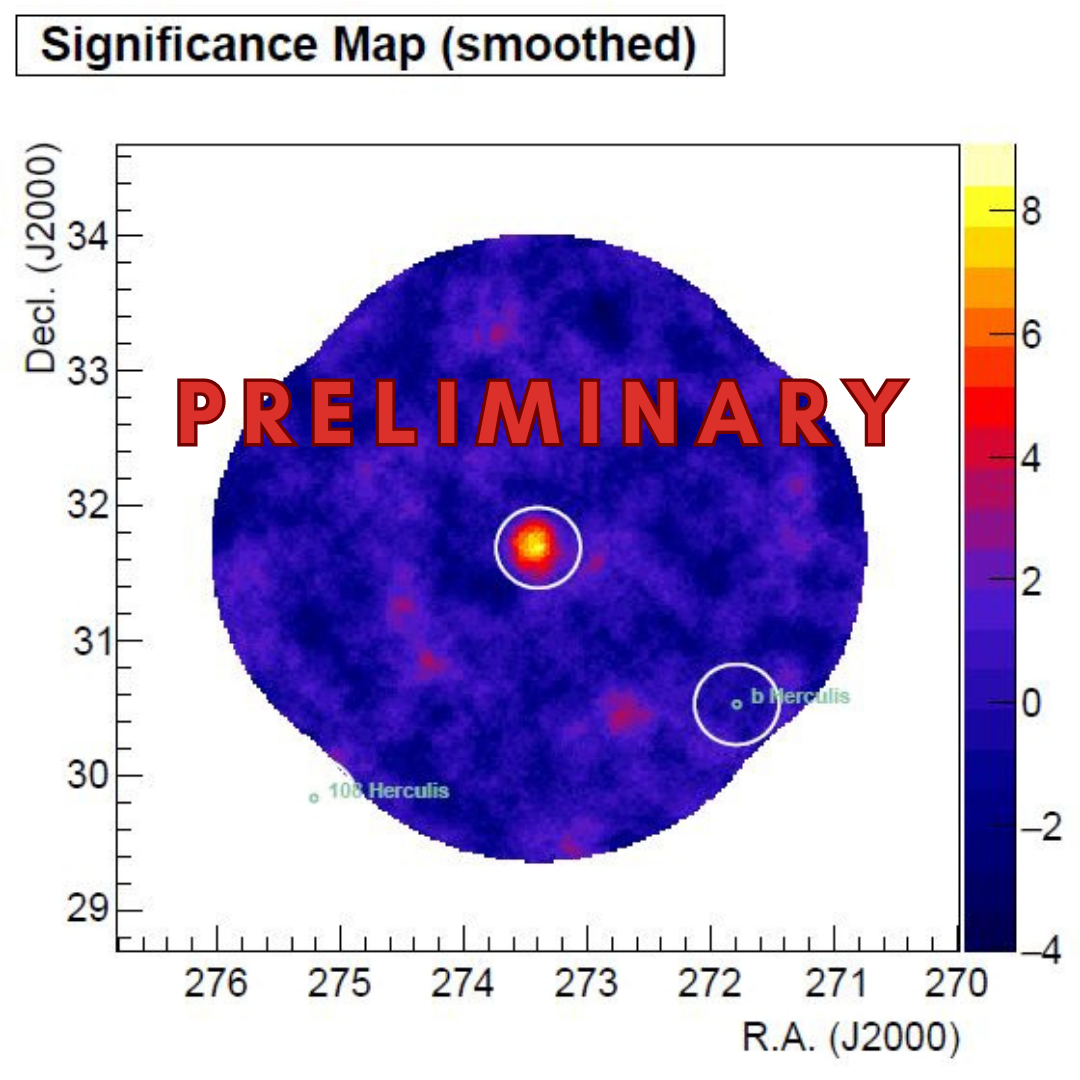}
  \caption{Smoothed significance sky map for the region of interest. The white circles indicate regions excluded from the background estimation, corresponding to known sources or bright stars.}
  \label{fig:significance}

\end{figure}

VERITAS devotes around half of its observation time to detect, follow up on, and monitor AGN sources \cite{Benbow2019ICRC}. VERITAS's AGN program thus allocates about 600 hours of good-weather time, each year to this task. One of the main focuses of the AGN program is the discovery and follow up observations of new VHE sources, comprising $\sim$40\% of AGN observation time. Most of these correspond to Target of Opportunity (ToO) observations triggered by any other multi-wavelength partner. The 2020 VERITAS B2 1811+31 monitoring is a relevant example of a successful ToO observation campaign. After a HE detection of enhanced activity by \textit{Fermi}-LAT and a 6$\sigma$ VHE detection by MAGIC, VERITAS started observing the source on October 15th, 2020. These observations spanned four consecutive nights, adding up to 4.5 good-quality hours. Measurements were performed using the standard “wobble” observation mode, with a 0.5\textdegree offset \cite{Fomin1994AP}. Our analysis of these observations yielded an 8.5$\sigma$ detection with an integral flux above the energy threshold of 200 GeV of $(1.74 \pm 0.36) \times 10^{-11} \, \textrm{cm}^{-2} \, \textrm{s}^{-1}$. Figure \ref{fig:significance} shows the corresponding significance map for B2 1811+31 during the 2020 flare.

\subsection{Fermi-LAT}
\textit{Fermi}-LAT is a large-area pair-conversion telescope aboard the \textit{Fermi} Gamma-ray Space Telescope. \textit{Fermi}-LAT is sensitive to photons within the energy range from 20 MeV to 300 GeV. \textit{Fermi}-LAT's main operational status is its survey mode, in which the LAT completes a comprehensive survey of the entire sky every 3 hours. Our analysis of data obtained by \textit{Fermi}-LAT was carried out using the \texttt{fermipy} \cite{Wood2017ICRC} Python package (version 1.1.4). We carried out two different temporal analyses of B2 1811+31: one that spanned the whole 12 years of LAT data (December 2008-December 2022), and one that focused on 2020 (January 2020-February 2021). Both analyses were carried out with a region of interest of 15$^\circ$ around the source, considering “source” class events (evclass=128) from both the front and back (evtype=3), and with energies between 100 MeV and 300 GeV. Binned likelihood analyses were performed adopting the 4FGL catalog \cite{Fermi2020} specifications for sources in the region of interest. Our fit freed the spectral parameters of all sources within 5$^\circ$ of B2 1811+31, and of all sources with TS$\geq$5 in the region of interest. The normalization of the isotropic and galactic diffuse components were also fit as free parameters. B2 1811+31 was significantly detected by \textit{Fermi}-LAT in the full dataset analysis, with a TS of 1603.36 ($\sqrt{TS} \sim \sigma$), assuming a power-law model. Similarly, for the 2020 analysis, a TS of 1767.12 was found.

\section{Light Curve Analysis}
Light curves were computed for both temporal analyses of \textit{Fermi}-LAT data, in order to identify a flaring period, and characterize the time evolution of said flare. For the light curve that employed B2 1811+31's full dataset, we computed 14-day time bins, 359 in total. In the case of the 2020 light curve, we employed 2-day time bins, amounting to 201 bins in total.

All bins in both light curves were subjected to a validation technique, with the intention of removing those bins that hadn't properly converged. We first represented the ratio $\frac{Flux}{\Delta Flux}$ versus the significance ($\sqrt{\left|TS\right|}$) for each light curve bin. We expect a proportional relation between the ratio $\frac{Flux}{\Delta Flux}$ and the significance [\cite{Valverde2020}]. Individual data points that deviated from said proportional relation were identified as having TS$<$0.01. Most of these bins had negative TS, a further indication of a fit convergence problem. All these flagged bins were removed from the light curve. The remaining bins were analyzed using a Bayesian Blocks statistical method, setting the false positive rate $p_0$ to 0.0027 (the value equivalent to 5$\sigma$ using Equation 13 of Scargle et al. (2013)) \cite{Scargle2013ApJ}.

\begin{figure}[!tbp]
  \centering\includegraphics[width=\textwidth]{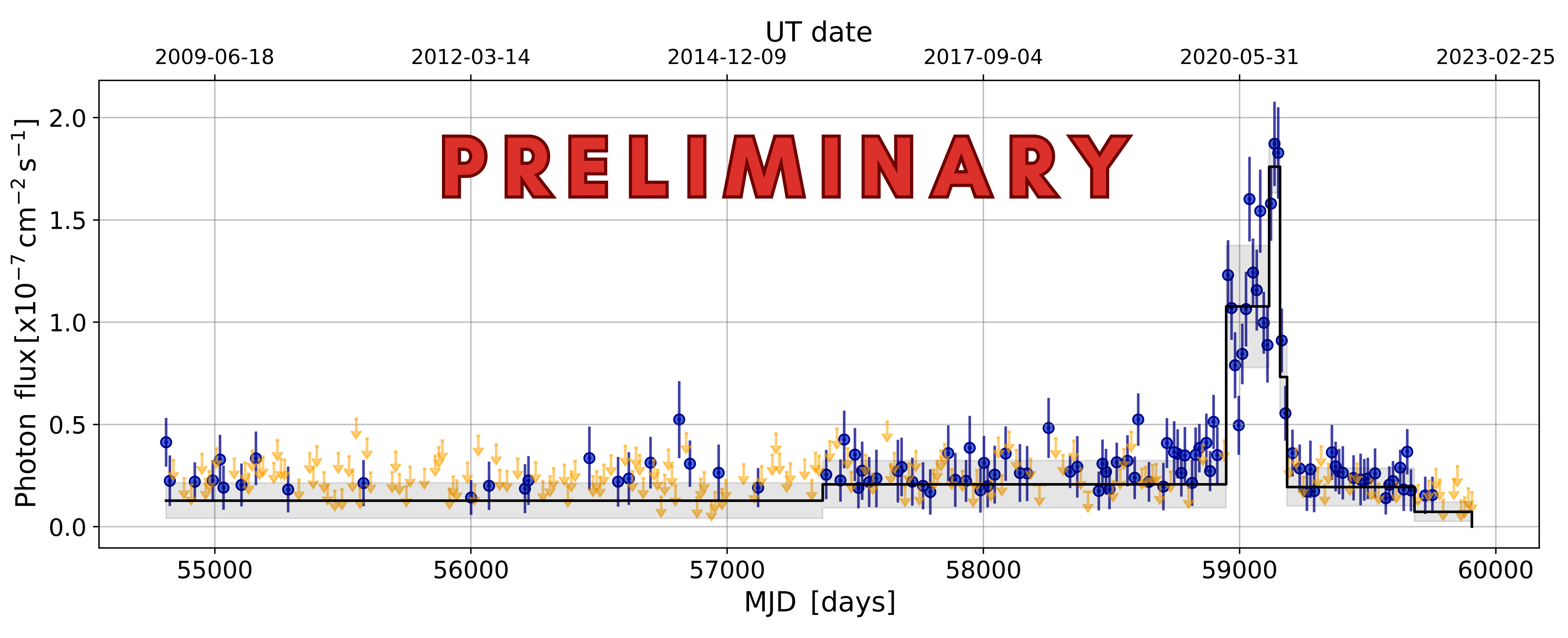}
  \caption{\textit{Fermi}-LAT B2 1811+31 12-year full dataset light curve, with 14-day bins. Orange arrows represent 95\% Upper Limits, for bins whose TS$<$4. Bins with TS$<$0.01 were excluded from the analysis. Black lines represent Bayesian Block fluxes, with gray shading marking the 1 standard deviation interval.}
  \label{fig:fulllc}
\end{figure}

We defined flares in two complementary ways, following the practice in Valverde (2020) \cite{Valverde2020}. In our first method, the data were first recursively fit to a constant function, initially the mean flux of all light curve data points. We defined the quiescent state as points that did not deviate from the mean flux more than 3$\sigma$, following equation 5.1 in Valverde (2020) \cite{Valverde2020}. We then recursively used the mean flux of points in the quiescent state as our constant function. We repeated this process until the quiescent state set of light curve points and the outlier set of points were fixed. If three consecutive bins were found in the outlier set, we considered that a flux. Our second method made use of the Bayesian Block analysis. For this method, we integrated the flare selection technique presented in Yoshida (2022) \cite{Yoshida2022ApJ}. We defined the quiescent flux as the Bayesian block with lowest flux that contained more data points that the mean number of data points per block. We established the flare threshold flux as the quiescent flux plus five times the mean flux uncertainty for all light curve bins (equation 1 in \cite{Yoshida2022ApJ}). All Bayesian blocks whose mean flux is above this threshold are considered flaring states.

When applying both of these methods to the full dataset light curve, represented in Figure \ref{fig:fulllc}, we find a complete agreement in terms of flare definition. Our second method identifies a flare spanning the third and four Bayesian Blocks, while our first method identifies all points in those two blocks, except for a single bin, as outliers, in flaring levels of flux. The fifth block is rejected from being part of the flare, but the first method finds that one out of the two bins comprised by this block is at abnormally high flux levels, showing a diffuse boundary regarding the end of the flare. An intention of better defining the flare limits prompted us to carry out a more detailed analysis of the 2020 flare with finer binning.

\begin{figure}[!tbp]
  \centering\includegraphics[width=\textwidth]{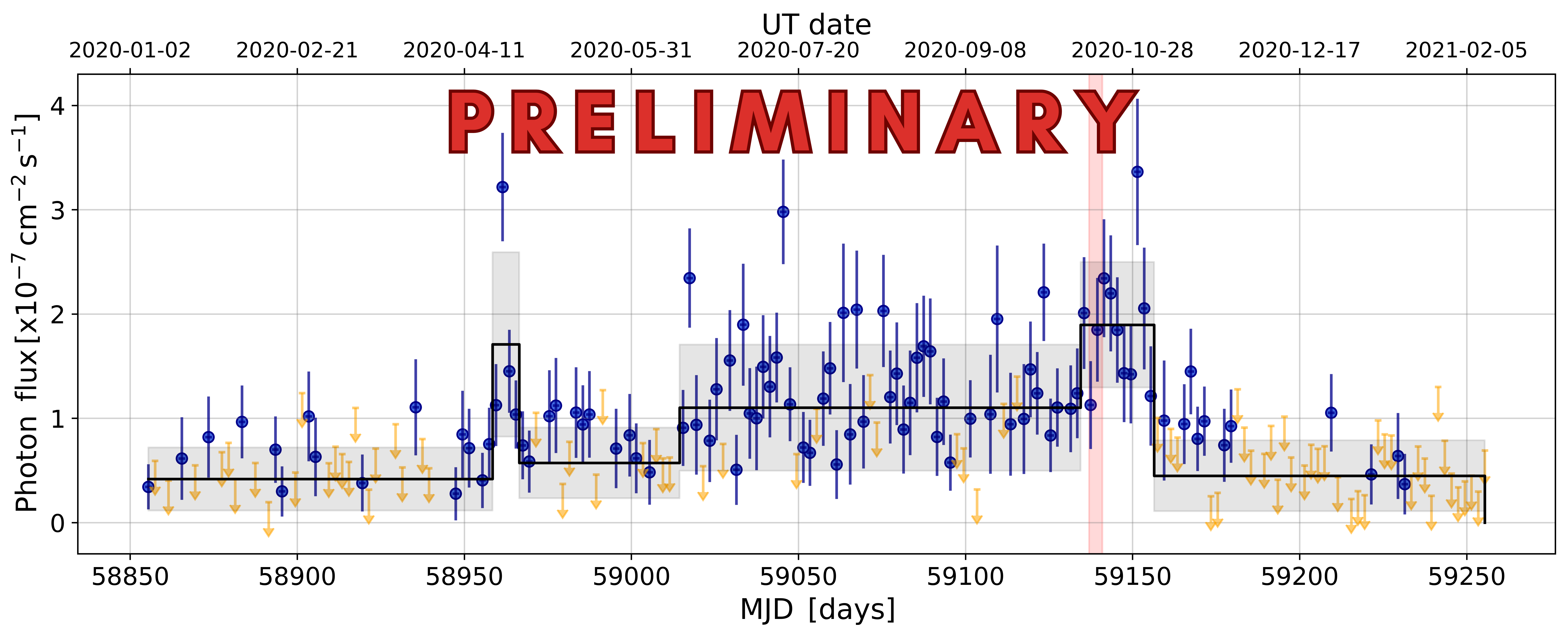}
  \caption{\textit{Fermi}-LAT B2 1811+31 2020 light curve, with 2-day bins. Orange arrows represent 95\% Upper Limits, for bins whose TS$<$4. Bins with TS$<$0.01 were excluded from the analysis. Black lines represent Bayesian Block fluxes, with gray shading marking the 1 Standard Deviation interval. Vertical light red shading indicates the time of VERITAS observations.}
  \label{fig:2020lc}
\end{figure}

Looking at Figure \ref{fig:2020lc}, we can identify a complex evolution within the 2020 flare, with no one clear exponential rise and decay. Applying our first method, we could define as the sole flare within this period the first five bins in the fifth Bayesian block, that are contained within the period in which VERITAS detected the source (shaded in red). However, no individual Bayesian block is detected as flaring following our second method. During various periods of elevated flux emissions (second, fourth and fifth Bayesian blocks in Figure \ref{fig:2020lc}), there are repeated instances of flux maxima occurring in isolated bins. This phenomenon points at a significant flux variability not being captured by the Bayesian Block analysis. A potential explanation for this discrepancy is that said variability could be occurring at daily timescales, sporadically throughout the six month flaring period. A closer inspection of these moments of extreme variability could inform us of mechanisms behind this flaring pattern. 

\section{Conclusions}
This work presents an analysis of the October 2020 VERITAS observations of B2 1811+31, along with a larger summary of the source's flaring period. The analysis of HE \textit{Fermi}-LAT frequencies show an interesting flare evolution that deviates from longer exponential rise and decay patterns. Instead, day-scale variability is observed, superimposed on the longer flare trends represented by the Bayesian Block analysis. In this context, it is especially interesting to note that the VERITAS detection of the source took place during a time of elevated flux in \textit{Fermi} frequencies, confirming the multiwavelength nature of the detected VHE flare. A further analysis of the source, tracing the development of the flare in different frequencies could help explain both its unusual variability and the presence of VHE emission. Such an analysis, along with a multiwavelength SED, will be presented in an upcoming publication.

\section*{Acknowledgments}
\label{sec:acknowledgements}

This research is supported by grants from the U.S. Department of Energy Office of Science, the U.S. National Science Foundation and the Smithsonian Institution, by NSERC in Canada, and by the Helmholtz Association in Germany. This research used resources provided by the Open Science Grid, which is supported by the National Science Foundation and the U.S. Department of Energy's Office of Science, and resources of the National Energy Research Scientific Computing Center (NERSC), a U.S. Department of Energy Office of Science User Facility operated under Contract No. DE-AC02-05CH11231. We acknowledge the excellent work of the technical support staff at the Fred Lawrence Whipple Observatory and at the collaborating institutions in the construction and operation of the instrument.

\clearpage

\section*{Full Author List: VERITAS Collaboration}

\scriptsize
\noindent
A.~Acharyya$^{1}$,
C.~B.~Adams$^{2}$,
A.~Archer$^{3}$,
P.~Bangale$^{4}$,
J.~T.~Bartkoske$^{5}$,
P.~Batista$^{6}$,
W.~Benbow$^{7}$,
J.~L.~Christiansen$^{8}$,
A.~J.~Chromey$^{7}$,
A.~Duerr$^{5}$,
M.~Errando$^{9}$,
Q.~Feng$^{7}$,
G.~M.~Foote$^{4}$,
L.~Fortson$^{10}$,
A.~Furniss$^{11, 12}$,
C.~Hahn$^{13}$,
W.~Hanlon$^{7}$,
O.~Hervet$^{12}$,
C.~E.~Hinrichs$^{7,14}$,
J.~Hoang$^{12}$,
J.~Holder$^{4}$,
Z.~Hughes$^{9}$,
T.~B.~Humensky$^{15,16}$,
W.~Jin$^{1}$,
M.~N.~Johnson$^{12}$,
M.~Kertzman$^{3}$,
M.~Kherlakian$^{6}$,
D.~Kieda$^{5}$,
T.~K.~Kleiner$^{6}$,
N.~Korzoun$^{4}$,
S.~Kumar$^{15}$,
M.~J.~Lang$^{17}$,
M.~Lundy$^{18}$,
G.~Maier$^{6}$,
C.~E~McGrath$^{19}$,
E.~T.~Meyer$^{13}$,
M.~J.~Millard$^{20}$,
C.~L.~Mooney$^{4}$,
P.~Moriarty$^{17}$,
R.~Mukherjee$^{21}$,
S.~O'Brien$^{18,22}$,
R.~A.~Ong$^{23}$,
N.~Park$^{24}$,
C.~Poggemann$^{8}$,
M.~Pohl$^{25,6}$,
E.~Pueschel$^{6}$,
J.~Quinn$^{19}$,
P.~L.~Rabinowitz$^{9}$,
K.~Ragan$^{18}$,
P.~T.~Reynolds$^{26}$,
D.~Ribeiro$^{10}$,
E.~Roache$^{7}$,
J.~L.~Ryan$^{23}$,
I.~Sadeh$^{6}$,
A.~C.~Sadun$^{27}$,
L.~Saha$^{7}$,
M.~Santander$^{1}$,
G.~H.~Sembroski$^{28}$,
R.~Shang$^{21}$,
M.~Splettstoesser$^{12}$,
A.~K.~Talluri$^{10}$,
J.~V.~Tucci$^{29}$,
V.~V.~Vassiliev$^{23}$,
A.~Weinstein$^{30}$,
D.~A.~Williams$^{12}$,
S.~L.~Wong$^{18}$,
and
J.~Woo$^{31}$\\
\\
\noindent
$^{1}${Department of Physics and Astronomy, University of Alabama, Tuscaloosa, AL 35487, USA}

\noindent
$^{2}${Physics Department, Columbia University, New York, NY 10027, USA}

\noindent
$^{3}${Department of Physics and Astronomy, DePauw University, Greencastle, IN 46135-0037, USA}

\noindent
$^{4}${Department of Physics and Astronomy and the Bartol Research Institute, University of Delaware, Newark, DE 19716, USA}

\noindent
$^{5}${Department of Physics and Astronomy, University of Utah, Salt Lake City, UT 84112, USA}

\noindent
$^{6}${DESY, Platanenallee 6, 15738 Zeuthen, Germany}

\noindent
$^{7}${Center for Astrophysics $|$ Harvard \& Smithsonian, Cambridge, MA 02138, USA}

\noindent
$^{8}${Physics Department, California Polytechnic State University, San Luis Obispo, CA 94307, USA}

\noindent
$^{9}${Department of Physics, Washington University, St. Louis, MO 63130, USA}

\noindent
$^{10}${School of Physics and Astronomy, University of Minnesota, Minneapolis, MN 55455, USA}

\noindent
$^{11}${Department of Physics, California State University - East Bay, Hayward, CA 94542, USA}

\noindent
$^{12}${Santa Cruz Institute for Particle Physics and Department of Physics, University of California, Santa Cruz, CA 95064, USA}

\noindent
$^{13}${Department of Physics, University of Maryland Baltimore County, 1000 Hilltop Circle, Baltimore, MD 21250, USA}

\noindent
$^{14}${Department of Physics and Astronomy, Dartmouth College, 6127 Wilder Laboratory, Hanover, NH 03755 USA}

\noindent
$^{15}${Department of Physics, University of Maryland, College Park, MD, USA }

\noindent
$^{16}${NASA GSFC, Greenbelt, MD 20771, USA}

\noindent
$^{17}${School of Natural Sciences, University of Galway, University Road, Galway, H91 TK33, Ireland}

\noindent
$^{18}${Physics Department, McGill University, Montreal, QC H3A 2T8, Canada}

\noindent
$^{19}${School of Physics, University College Dublin, Belfield, Dublin 4, Ireland}

\noindent
$^{20}${Department of Physics and Astronomy, University of Iowa, Van Allen Hall, Iowa City, IA 52242, USA}

\noindent
$^{21}${Department of Physics and Astronomy, Barnard College, Columbia University, NY 10027, USA}

\noindent
$^{22}${ Arthur B. McDonald Canadian Astroparticle Physics Research Institute, 64 Bader Lane, Queen's University, Kingston, ON Canada, K7L 3N6}

\noindent
$^{23}${Department of Physics and Astronomy, University of California, Los Angeles, CA 90095, USA}

\noindent
$^{24}${Department of Physics, Engineering Physics and Astronomy, Queen's University, Kingston, ON K7L 3N6, Canada}

\noindent
$^{25}${Institute of Physics and Astronomy, University of Potsdam, 14476 Potsdam-Golm, Germany}

\noindent
$^{26}${Department of Physical Sciences, Munster Technological University, Bishopstown, Cork, T12 P928, Ireland}

\noindent
$^{27}${Department of Physics, University of Colorado Denver, Denver, Colorado, CO 80217, USA}
% $^{27}${Department of Physics, University of Colorado Denver, Campus Box 157, P.O.Box 173364, Denver CO 80217, USA }

\noindent
$^{28}${Department of Physics and Astronomy, Purdue University, West Lafayette, IN 47907, USA}

\noindent
$^{29}${Department of Physics, Indiana University-Purdue University Indianapolis, Indianapolis, IN 46202, USA}

\noindent
$^{30}${Department of Physics and Astronomy, Iowa State University, Ames, IA 50011, USA}

\noindent
$^{31}${Columbia Astrophysics Laboratory, Columbia University, New York, NY 10027, USA}

\end{document}